\documentclass{jais}       

\journal{JAIS-ID}
\vol{xx}

\received{14 August 2026}
\published{}

\usepackage{amsmath}
\usepackage{amssymb}
\usepackage[hidelinks]{hyperref}
\usepackage{tabularx}
\usepackage{graphicx}
\usepackage{dcolumn}
\usepackage{subcaption}
\usepackage{multirow}
\usepackage{booktabs}

\def\be{\begin{equation}}
\def\ee{\end{equation}}
\def\bea{\begin{eqnarray}}
\def\eea{\end{eqnarray}}

\renewcommand{\thefootnote}{\fnsymbol{footnote}}

\begin{document}

\title{From Simulation to Real Scans: Anomaly Detection in Maritime Cargo with Muon Scattering Tomography}

\author{
Angel Bueno Rodriguez\auno{1},
Christina Hrytsiuk\auno{2}, 
Maximilian Perez Prada\auno{1},
Kaarel Tark\auno{2}, 
Felix Sattler\auno{1},
Jean Marco Alameddine\auno{1}, 
Madis Kiisk\auno{2},
Maurice Stephan\auno{1}, 
and Sarah Barnes\auno{1}
}

\address{
\begin{flushleft}
    
$^1$German Aerospace Center, Institute for the Protection of Maritime Infrastructures, Bremerhaven, Germany\\
$^2$GScan O\"{U}, M\"{a}ealuse 2/1, Tallinn, Estonia\\
\end{flushleft}
}

\begin{abstract}
\begin{flushleft}
\begin{minipage}{0.99\textwidth}
Maritime cargo inspection requires imaging technologies capable of detecting concealed threats within dense, sealed containers, a role for which Muon Scattering Tomography (MST)
is well suited: it images their interior through the density-dependent deflection of
naturally occurring cosmic muons. However, MST remains constrained by the scarcity of labeled scans and by a cosmic muon flux that is both low and stochastic. Anomaly detection algorithms must therefore be trained on simulations, yet operate on measured scans acquired under different conditions, a sim-to-real gap that remains a central obstacle to operational deployment. We present the first end-to-end anomaly detection framework for maritime MST, from physically consistent simulations to validation on real container scans from the SilentBorder demonstration campaign. The task is cast as an out-of-distribution problem: the framework learns the spatial configurations of benign cargo and flags threats as deviations in the reconstruction error space, remaining agnostic to threat type and geometry. An attention U-Net, trained exclusively on benign synthetic scenes, preserves small-scale scattering signatures through its skip connections, and contraband consequently persists in the pixel-wise reconstruction error instead of being absorbed into the reconstructed background. A scoring function, the Homogeneity Index (HI), suppresses spatially uniform cosmic-ray statistical noise while amplifying coherent anomaly signatures: where pixel-level metrics collapse under a change of cargo configuration, HI retains its discriminative power. We evaluate three training strategies across two distinct cargo configurations under operational one-hour scan times, and test the best model on real muon cargo scans. The results for the studied scenarios indicate that the sim-to-real gap can be bridged. 
\end{minipage}

\par\vspace{8pt}

\par\vspace{4pt}

\noindent\textit{DOI:} 10.31526/JAIS.2026.ID

\end{flushleft}

\end{abstract}

\maketitle

\renewcommand{\thefootnote}{\fnsymbol{footnote}}
\footnotetext{Corresponding author: angel.bueno@dlr.de. Submitted to the Journal of
Advanced Instrumentation in Science (JAIS), Muographers 2026 proceedings.}

\section{Introduction}

Global maritime trade continues to grow, with major shipping routes expanding~\cite{TAGAWA2025104376} and millions of containers handled annually at seaports worldwide. This creates a demand for robust, non-intrusive inspection (NII) technologies capable of detecting illicit materials within shipping containers. Traditional NII systems, primarily based on high-energy X-ray or gamma-ray sources, face significant limitations, from operational safety to logistical constraints, when dealing with heavily shielded cargo~\cite{sarah2023}. Muon Scattering Tomography (MST) has emerged as a suitable alternative, leveraging the high penetrative power of naturally occurring cosmic-ray muons, subatomic particles generated in the upper atmosphere, to provide a passive imaging modality~\cite{bonechiyandrea}. MST imaging relies on multiple Coulomb scattering: as muons traverse matter, their trajectory is deflected by an angle whose distribution has a width $\theta_0$, as described by: 

\begin{equation}
    \theta_0 = \frac{13.6 \text{ MeV}}{\beta c\, p_\mu} \sqrt{\frac{\ell}{X_0}} \left[ 1 + 0.038 \ln \left( \frac{\ell}{X_0} \right) \right]
\label{eq:msc}
\end{equation}

Here, $\ell$ is the thickness of material traversed, $p_\mu$ is the muon momentum, $\beta c$ is its velocity, and $X_0$ is the radiation length of a material~\cite{borozdin2003radiographic}. Since $X_0$ is inversely related to the atomic number $Z$, materials with high $Z$, such as uranium, iron, or lead, cause larger scattering than low-$Z$ materials, such as water or wood. By tracking the incoming and outgoing trajectories of individual muons, these scattering deflections are aggregated to reconstruct a scattering-density volume that reveals the internal structure of the cargo.\\

In operational settings, however, MST faces practical challenges. The cosmic muon flux is low for imaging purposes (approximately $1 \text{ cm}^{-2} \text{ min}^{-1}$) and inherently stochastic. Hence, the scattering-density volumes reconstructed under the short acquisition times of port operations are sparse and noisy. Recent research efforts, such as the SilentBorder project\footnote{SilentBorder is an EU Horizon~2020 project developing a cosmic-muon tomography scanner for cargo border inspection.}, address these hurdles with advanced tomographic reconstruction; yet interpreting the resulting scattering-density volumes remains difficult~\cite{borozdin2023methods, odonnell2025upsampling, riggi2013muon}, as maritime cargo is structurally diverse and real anomalous measurements are too scarce to train supervised recognition systems. Moreover, the automated interpretation of MST volumes has so far been studied almost exclusively in simulations~\cite{georgadze2025illicit,sattler2025framework}, leaving an open question as to whether such methods hold on real container measurements, that is, whether the simulation to real (sim-to-real) gap can be bridged.\\

We therefore introduce an unsupervised anomaly detection framework that learns the spatial configurations of benign cargo from physically consistent synthetic data and flags deviations as out-of-distribution (OOD). The study focuses on Intermediate Bulk Containers (IBCs), one of the most common packaging formats in maritime transport, and examines whether a model trained on a single IBC configuration (IBC1) transfers to a dual IBC configuration (IBC2), that is, whether a single model suffices for both. In an operational port, the content and arrangement of the inspection volume change from scan to scan, and a detection system cannot be retrained for every new configuration. The simulated IBC1-to-IBC2 transition therefore acts as a controlled proxy for that variability, and
the degree of adaptation between configurations is a direct measure of operational viability. We additionally study a Cold-start strategy, in which the IBC1-pretrained model is fine-tuned on a small set of synthetic IBC2 scenes, to determine whether the pretrained representation transfers to the new cargo configuration. A further challenge is that reconstruction artifacts and the stochastic cosmic-ray background yield errors that standard pixel-level scores misinterpret as anomalies. To address this, we introduce a spatially aware scoring function that separates coherent, object-like deviations from artifact-driven noise, making the anomaly decision rule robust to the noise floor of each MST scan. The framework is validated on real muon scans of cargo acquired before and after the SilentBorder demonstration, without access to contraband labels at any stage of training. The main contributions of this study are:

\begin{enumerate}

    \item \textbf{Operational sim-to-real validation}: To our knowledge, 
    the first end-to-end anomaly detection framework validated on real muon-scattering measurements of maritime cargo, acquired through one-hour scans during the pre- and post-SilentBorder demonstration.
    
    \item \textbf{Physically consistent synthetic data generation}: A synthetic dataset of benign and anomalous cargo scenarios (single- and dual-IBC configurations), compliant with EU transport regulations and replicating the setup of the SilentBorder campaign, enabling
    the study of anomaly detection transferability to real measurements.

    \item \textbf{Spatially aware anomaly scoring}: The Homogeneity Index (HI), a scoring function that detects whether reconstruction errors are spatially localized rather than spread uniformly across the map. HI outperforms pixel-level scoring and enables transfer both between cargo configurations and from simulated to real measurement data.
\end{enumerate}

\section{Anomaly Detection in MST}
\label{sec:anomalydet}

In maritime security, characterizing contraband is very challenging due to the vast diversity of illicit cargo configurations. We therefore model the baseline of benign cargo and treat any deviation from it as a candidate threat. However, the reconstructed scattering-density volume is a difficult input for anomaly detection algorithms. As Equation~(\ref{eq:msc}) shows, the scattering width $\theta_0$ depends jointly on the traversed material ($\ell/X_0$) and on the muon momentum $p_\mu$: without measuring individual momenta, a deflection produced by a low-momentum muon in a light material is kinematically indistinguishable from that of a high-momentum muon in a dense one~\cite{bueno2025pillar}. Machine-learning approaches to estimating the muon momentum from scattering data are being explored~\cite{bury2025momentum}, but no such information is available in the detectors considered here. In addition, the low-count muon statistics introduce stochastic fluctuations across the entire voxel grid that mask localized anomalies. To mitigate the latter, we work on 2D sections of the reconstructed volume rather than on the full 3D grid. The reconstruction stage outputs a scattering-density volume $V \in \mathbb{R}^{N_X \times N_Y \times N_Z}$, where each voxel encodes the accumulated scattering density at its spatial location. The volume is sectioned along the vertical axis into a set of 2D axial scattering maps: each slice $\mathbf{x}_z = V(\cdot, \cdot, z) \in \mathbb{R}^{N_X \times N_Y}$ is the $(X, Y)$ matrix corresponding to height $z$, analogous to slice-based analysis in volumetric medical imaging~\cite{medicalref}. Combined with local averaging across neighboring slices (Section~\ref{sec:expmen}), 
this sectioning suppresses spatially uncorrelated noise fluctuations and amplifies persistent scattering signatures from dense objects, at the cost of depth resolution along the sectioning axis.

Let $\mathcal{X}_{\mathrm{norm}}$ denote the space of 2D axial scattering maps of physically consistent, benign cargo, and $p_{\mathrm{norm}}(\mathbf{x})$ the probability density of these maps. The objective is to characterize this normal data implicitly, without estimating its density. Rather than modeling $p_{\mathrm{norm}}(\mathbf{x})$ directly, we learn to reconstruct samples drawn from $\mathcal{X}_{\mathrm{norm}}$ and use the reconstruction error as evidence of anomaly. During deployment, any reconstructed 2D scattering map $\mathbf{x}_{\mathrm{test}}$ 
is assigned an anomaly score $s(\mathbf{x}_{\mathrm{test}})$, defined in Sections~\ref{sec:score} and~\ref{sec:hi}; if $s(\mathbf{x}_{\mathrm{test}}) > \tau$, for an operational threshold $\tau$, the sample is flagged as OOD and triggers an anomaly alert. This inverts the detection problem: the system does not need to know what an illicit good looks like; it only needs to recognize that a scattering signature does not belong in a benign cargo scan.

\begin{figure*}[htbp!]
    \centering
    \includegraphics[width=0.99\linewidth]{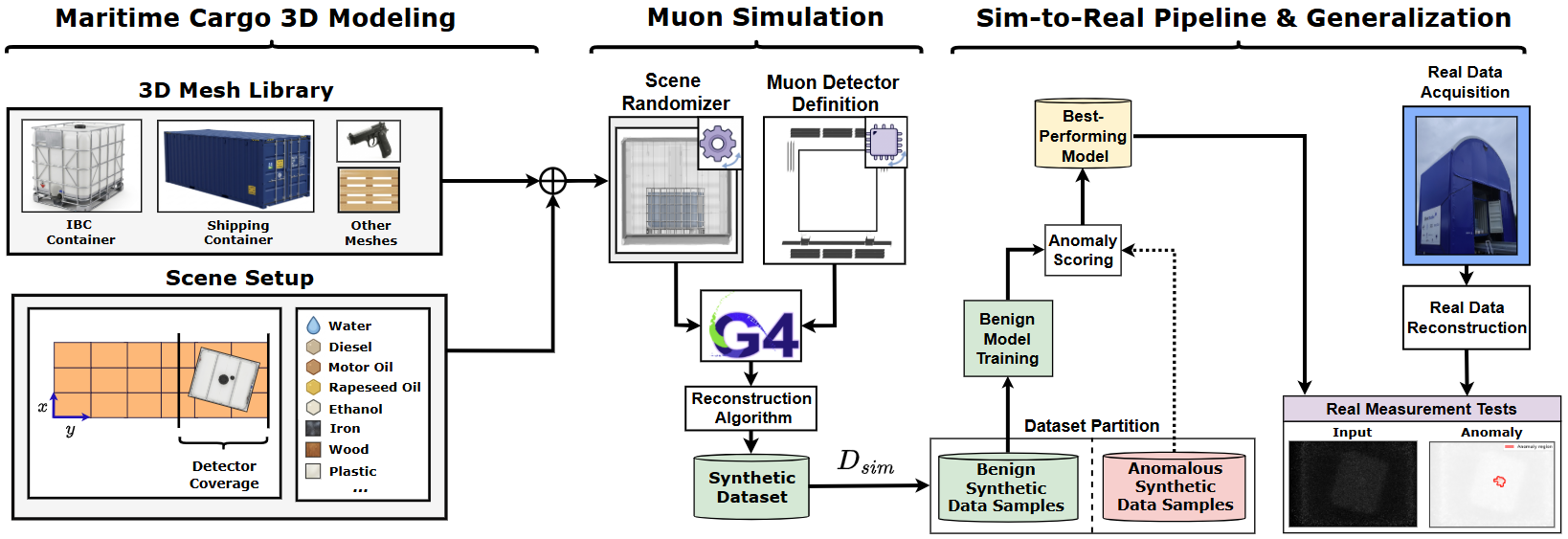}
    \caption{Overview of the end-to-end framework. \textbf{Left}: the mesh library and material palette from which the IBC scenes are assembled. \textbf{Centre}: each randomized scene is simulated in Geant4 and reconstructed into a scattering-density volume, yielding the synthetic dataset $D_{sim}$. \textbf{Right}: $D_{sim}$ is partitioned into benign and anomalous samples. The attention U-Net is trained only on the benign partition; the anomalous partition serves exclusively to validate the anomaly scores and never enters training (dashed arrow). The best-performing checkpoint is then applied without retraining to the real measurement campaign, whose scans are reconstructed with the same algorithm and resampled to the network input size. The red contour marks the anomaly region returned by the scoring stage.}
  \label{fig:figura1}
\end{figure*}

\subsection{Preserving Spatial Resolution with U-Net}

The reconstruction task is cast as a denoising procedure. We model it as an operator $f_\theta : \mathbb{R}^{N_X \times N_Y} \rightarrow \mathbb{R}^{N_X \times N_Y}$ mapping an input axial map $\mathbf{x}$ to its denoised estimate $\hat{\mathbf{x}} = f_\theta(\mathbf{x})$. Trained exclusively on $\mathcal{X}_{\mathrm{norm}}$, $f_\theta$ becomes an implicit model of benign cargo: inputs drawn from it are reconstructed with high fidelity, while OOD scattering signatures yield large reconstruction errors. A conventional convolutional autoencoder is not well suited here, since its bottleneck
discards the fine spatial detail in which the anomaly signature resides. We adopt a U-Net backbone for $f_\theta$~\cite{UNet}, an encoder--decoder network whose skip connections reinject high-resolution encoder features into the decoder; attention gates modulate these connections at each up-sampling stage, suppressing features in regions inconsistent with the learned normality.

\subsection{Anomaly Scoring via Reconstruction Error}
\label{sec:score}

The parameters $\theta$ of the operator $f_\theta$ are optimized by minimizing a composite reconstruction loss $\mathcal{L}$ over the samples in $\mathcal{X}_{\mathrm{norm}}$:

\begin{equation}
\begin{split}
\mathcal{L} \;=\;\; & 
\underbrace{\alpha \cdot \mathrm{MSE}\!\left(\mathbf{x},\, f_\theta(\mathbf{x})\right)}_{\text{pixel fidelity}}
\;+\; 
\underbrace{(1-\alpha)\cdot\Big(1 - \mathrm{SSIM}\!\left(\mathbf{x},\, f_\theta(\mathbf{x})\right)\Big)}_{\text{perceptual similarity}} \\[6pt]
& +\; 
\underbrace{\lambda_{TV}\cdot \mathrm{TV}\!\left(f_\theta(\mathbf{x})\right)}_{\text{smoothness}}
\end{split}
\label{eq:2}
\end{equation}

\noindent The pixel fidelity term, the MSE (Mean Squared Error) with $\alpha = 0.9$, measures the pixel-level deviation between the input and the reconstruction, quantifying the divergence of the network's output from the learned scattering background. The perceptual similarity term, the SSIM (Structural Similarity Index)~\cite{SSIM}, complements pixel fidelity by enforcing structural coherence: it penalizes reconstructions that fail to preserve edges, contrast gradients, and local texture patterns. The smoothness term, the TV (total variation) of the reconstruction ($\lambda_{TV} =10^{-3}$), acts as a regularizer that prevents the model from fitting stochastic muon noise as if it were real structure.\\

At inference, the residual map $E = |\mathbf{x} - f_\theta(\mathbf{x})|$ is the difference between the input and its denoised estimate, that is, the component the network attributes to noise. On benign cargo, this component is non-zero across the whole map, whereas structures absent from $\mathcal{X}_{\mathrm{norm}}$, such as a submerged weapon occupying only a few pixels, cannot be reproduced by $f_\theta$ and emerge as spatially compact spikes on top of it. A uniform noise floor is therefore present by construction rather than being incidental, which makes converting this pixel-level residual into a robust scalar anomaly score non-trivial: simple aggregation statistics such as MAE (Mean Absolute Error) reduce the residual map to a single global value dominated by that floor, masking the localized anomaly signal. Since this noise floor varies across cargo configurations, pixel-level scores collapse entirely in cross-domain evaluation (Section~\ref{sec:results}). Section~\ref{sec:hi} introduces a scoring function designed to be insensitive to it.

\subsection{Spatially Aware Scoring via Homogeneity Index}
\label{sec:hi}
To overcome the limitations of pixel-level scores, we introduce a cell-based scoring function, the Homogeneity Index (HI). Given a residual map $E$ of dimensions $H \times W$, we partition it into a regular grid of $n$ non-overlapping cells of size $c \times c$ pixels. For the full map, a reference probability distribution $p_i$ is computed over $k$ intensity bins; for each cell $j$, the local probability $p_{ij}$ is computed over the same bins. The HI is then defined as:

\begin{equation}
\mathrm{HI} = \sum_{i=1}^{k} \sqrt{ \frac{1}{n} \sum_{j=1}^{n} (p_{ij} - p_i)^2 }
\label{eq:hi}
\end{equation}

\noindent A homogeneous residual map, one with spatially uniform noise 
and no localized anomaly, produces similar distributions across all 
cells, yielding $\mathrm{HI} \approx 0$. When a spatially localized object, such  as a weapon, creates a cluster of high-error pixels concentrated in a few  cells, those cells exhibit a distributional shift relative to the global histogram, yielding a high HI score.\\

The cell size $c$ controls the spatial resolution of the scoring function. We evaluate two variants: HI-32 ($c = 32$) and HI-16 ($c = 16$); the finer grid better resolves small objects that occupy only a small fraction of the axial scattering map. The grid is truncated to $\lfloor H/c \rfloor \times \lfloor W/c \rfloor$ cells, so that for the $195 \times 290$ maps, HI-32 uses $n = 54$ cells and HI-16 uses $n = 216$, both covering the same central $192 \times 288$ region. Histograms are computed over $k = 16$ intensity bins. Each residual map is min-max normalized to $[0,1]$ before binning, so that HI measures relative distributional deviations within each map rather than absolute error magnitudes, which makes it largely insensitive to the global noise floor.

\begin{figure}[t]
\centering
\includegraphics[width=0.50\columnwidth, angle=0]{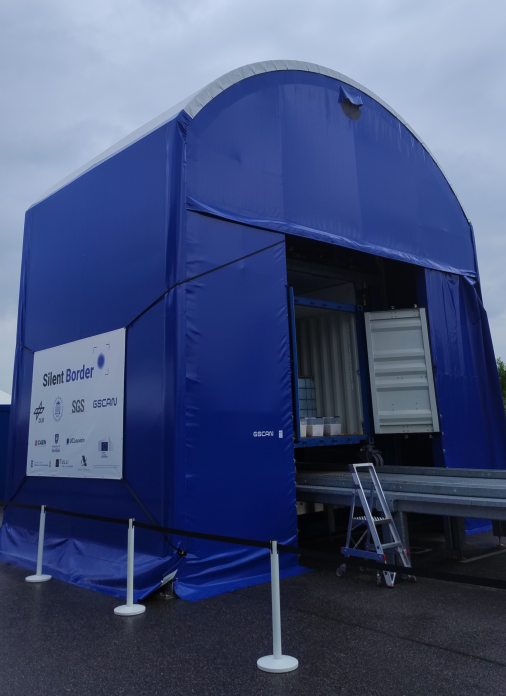}
\caption{The SilentBorder muon scanner during the preparations for the measurement campaigns. The container remained sealed throughout all acquisitions. Source: A. Bueno.}
\label{fig:setup}
\end{figure}

\section{Operational Setup}
\label{sec:operation-setup}
Modeling benign cargo configurations ($\mathcal{X}_{\mathrm{norm}}$) requires a dataset that real-world container scans cannot supply in the required diversity. A synthetic data generation pipeline was therefore developed to replicate real container scans. Figure~\ref{fig:figura1} presents an overview of the complete framework, from scene
definition to the evaluation on real scans. The data generation stages are detailed below:\\

\noindent\textbf{Synthetic data creation:} Operational anomaly detection in maritime transport must account for bulk liquids, among the most common cargo types shipped in IBCs. Two cargo configurations are considered, matching the real measurement campaigns: \textbf{IBC1}, with 
a single IBC in the inspection volume, and \textbf{IBC2}, with two IBCs placed side by side. For both cargo configurations, the Blender-to-Geant4 (B2G4) framework~\cite{rodriguez2024b2g4} was used to render physically consistent scenarios in compliance with EU transport regulations~\cite{eutransportIBC2021}, with liquid compositions representing standard IBC transport cargo: ethanol ($\rho = 0.789$~g/cm$^3$), diesel ($\rho = 0.82$~g/cm$^3$), motor oil ($\rho = 0.88$~g/cm$^3$), rapeseed oil ($\rho = 0.91$~g/cm$^3$), and water ($\rho = 1.00$~g/cm$^3$). Benign scenes contain only the liquid-filled IBCs; anomalous scenes additionally conceal handgun replicas with metallic slides and polymer frames submerged in the liquids. To prevent the network from memorizing fixed spatial coordinates, randomizations were applied: spatial jittering, multi-axis rotational perturbations ($\pm 2^{\circ}$), and varying submersion depths, bounded by the dimensional constraints of commercial IBCs and reflecting field-relevant concealment practices. For each scene, the set of slice indices where the contraband intersects the axial plane is retained as ground truth for slice- and scene-level evaluation.\\

\noindent\textbf{Simulation:} The detection system consists of eight hodoscopes, each measuring $2047.5 \times 1023.5$~mm$^2$ in active area. Each hodoscope is organized into groups of three plastic-scintillator-based plates, each $0.9$~mm thick, on all sides of the inspection volume (top, bottom, left, and right), resulting in a total of 24 detector planes (Figure~\ref{fig:figura1}). This arrangement provides angular coverage on all four sides and enables tracking of both incoming and outgoing muon trajectories throughout the cargo volume. The cosmic-ray source was modeled using the EcoMug generator~\cite{pagano2021ecomug}, which features a hemisphere-shaped emission surface with a 4~m radius, implemented in Geant4 v11.0.3~\cite{agostinelli2003geant4}. The detector configuration was developed within the SilentBorder project~\cite{zaher2025optimization}. \\

\noindent\textbf{Reconstruction method:} The scattering-density volumes used as input to the anomaly detection pipeline were generated by GScan O\"{U} using a proprietary reconstruction 
methodology~\cite{kiisk2026detector, georgadze2023method}, based on an optimized back-projection algorithm embedded in the SilentBorder detector hardware. From the perspective of the anomaly detection framework, the reconstruction stage is a fixed black box: volumes are received \textit{as-is}, with no access to raw detector hits or the reconstruction internals, and the same algorithm processes both synthetic and real measurements at a known voxel resolution of 1~cm. \\

\noindent\textbf{Dataset partitioning:} The IBC1 dataset comprises 147 synthetic training
scenes, with 29 held-out benign test scenes and 20 anomalous test scenes. The IBC2 dataset comprises 35 synthetic training scenes, with 7 held-out benign test scenes and 20 anomalous test scenes.\\

\noindent\textbf{Real measurement campaign:} To validate the sim-to-real transferability, dedicated measurement campaigns were conducted before and after the SilentBorder demonstration using the operational muon detector developed within the consortium. Figure~\ref{fig:setup} shows the scanning system; the container remained sealed throughout all acquisitions. The measured scenes replicate the synthetic configurations, with single (IBC1) and dual (IBC2) containers and handguns as contraband; for logistical reasons, a single bundle of three tightly packed handguns was used, manually placed within the IBCs. All measured scenes contain contraband, as no benign real acquisition was available; the real-data evidence therefore establishes detection and correct localization under operational conditions, while the metrics are quantified on synthetic data only. 

\begin{table}[htpb!]
\centering
\caption{Training strategies and best validation PSNR on benign synthetic cargo.}
\label{tab:training}
\begin{tabular}{c c c c}
\toprule
\textbf{Model} & \textbf{Train Data} & \textbf{Init.} & 
\textbf{Best PSNR} \\
\midrule
IBC1-only  & 147 IBC1 & Scratch    & 31.71~dB \\
IBC2-only  & 35 IBC2 & Scratch    & 30.14~dB \\
Cold-start & 35 IBC2 & IBC1 ckpt. & 31.50~dB \\
\bottomrule
\end{tabular}
\end{table}

\begin{table*}[htpb!]
\centering
\caption{Scene-level performance metrics across all experimental setups.}
\label{tab:scene_metrics}
\begin{tabular}{c c cc cc}
\toprule
\multirow{2}{*}{\textbf{Model}} & \multirow{2}{*}{\textbf{Scoring}} &
\multicolumn{2}{c}{\textbf{IBC1 test}} &
\multicolumn{2}{c}{\textbf{IBC2 test}} \\
\cmidrule(lr){3-4} \cmidrule(lr){5-6}
& & \textbf{AUROC~$\uparrow$} & \textbf{AUPRC~$\uparrow$} &
\textbf{AUROC~$\uparrow$} & \textbf{AUPRC~$\uparrow$} \\
\midrule
\multirow{3}{*}{IBC1-only}
& MAE     & 0.983 & 0.978 & 0.379 & 0.670 \\
& HI-32   & 0.898 & 0.819 & 0.743 & 0.919 \\
& HI-16   & 0.922 & 0.852 & 0.671 & 0.888 \\
\midrule
\multirow{3}{*}{IBC2-only}
& MAE     & 0.438 & 0.358 & 0.593 & 0.825 \\
& HI-32   & 0.897 & 0.811 & 0.750 & 0.916 \\
& HI-16   & 0.926 & 0.859 & 0.736 & 0.914 \\
\midrule
\multirow{3}{*}{Cold-start}
& MAE     & 0.988 & 0.984 & 0.429 & 0.695 \\
& HI-32   & 0.898 & 0.813 & 0.750 & 0.919 \\
& HI-16   & 0.922 & 0.852 & 0.679 & 0.894 \\
\bottomrule
\end{tabular}
\end{table*}

\section{Experimental Methodology}
\label{sec:expmen}

\subsection{Data Pre-processing}

\noindent\textbf{Z-slicing:} Each 3D volumetric scan is sectioned into 2D axial scattering maps, as described in Section~\ref{sec:anomalydet}, retaining the Z-slices in $[71, 328]$, for a total of 258 slices per scene, and cropping each
map to $195 \times 290$ pixels to exclude low-signal border regions. The reconstructed volume covers the full inspection region enclosed by the hodoscopes rather than the entire container: at the 1~cm voxel resolution, the retained range in $Z$ spans 2.58~m along the vertical axis and the crop corresponds to a horizontal section of $1.95 \times 2.90$~m$^2$, with $X$ across the container width and $Y$ along its length. The crop thus retains the IBC footprint together with a margin of empty inspection volume that supplies the benign background context.\\

\noindent\textbf{Synthetic data processing:} Each slice undergoes three data pre-processing steps. First, a Gaussian blur ($3\times3$ kernel, $\sigma = 0.25$) reduces high-frequency reconstruction artifacts. Next, a four-level discrete wavelet transform is applied, combining soft thresholding~\cite{donoho1995} for noise suppression with per-subband gain factors for structure enhancement, following a multiscale enhancement scheme~\cite{zong1998}. Gains of 1.8 (coarse subbands) and 2.5 (fine subbands) and a soft threshold of 0.05 were determined empirically on benign synthetic scenes; no anomalous or real scan was used to select them. A sliding-window average of three consecutive Z-slices replaces each input slice with the mean of itself and its Z-neighbors. Together, these steps amplify coherent scattering structures and suppress stochastic noise. Ground truth labels are defined as in Section~\ref{sec:operation-setup}.\\

\noindent\textbf{Real data processing}: Real volumes are resampled and cropped to the $195 \times 290$ input size of the
network, and are otherwise passed to the model as received from the reconstruction stage. Therefore, real scans reach the network with a noise floor that differs from anything seen during training, on top of the domain gap between simulated and measured scattering. No real scan was used to select any pre-processing or training parameter.

\subsection{Optimization Procedure}

The U-Net backbone is optimized to denoise pre-processed 2D scattering maps of benign cargo by minimizing the loss of Equation~(\ref{eq:2}). This U-Net follows an attention-gated architecture with five encoder levels (16, 32, 64, 128, and 256 channels). Table~\ref{tab:training} summarizes the three training strategies. IBC1-only and IBC2-only are trained from scratch on their respective configurations with a learning rate of $10^{-3}$; the third strategy, Cold-start, initializes from the IBC1-only checkpoint and fine-tunes on the IBC2 training set with a learning rate of $2\times10^{-4}$, so that the representation acquired during pretraining is preserved. All models use AdamW with a weight decay of $10^{-4}$ and ReduceLROnPlateau scheduling (factor 0.5, patience 5 epochs); early stopping triggers if the validation PSNR does not improve by more than 0.05~dB over eight consecutive epochs. The resulting checkpoints, selected on the reconstruction quality (PSNR) of benign scenes alone, are used for all anomaly detection inferences.

\subsection{Performance Metrics}

Scoring and performance are evaluated at both the slice and scene levels, with a twofold aim: to establish the feasibility of anomaly detection under operational port conditions, and to show that the spatial structure of the residual is the property that makes detection transferable across cargo configurations. Three scoring functions are compared: MAE, the mean of the residual map $E = |\mathbf{x} - f_\theta(\mathbf{x})|$, and HI with cell sizes $c = 32$ and $c = 16$, as defined in Section~\ref{sec:hi}. Comparing MAE and HI isolates the impact of spatial awareness: since the residual map $E$ is identical in both cases, any improvement from MAE to HI is attributable solely to the cell-based distributional scoring.\\

\noindent \textbf{Scene level}: The anomaly score of each scene is the mean of its
slice-level scores. We report the Area Under the Receiver Operating Characteristic Curve (AUROC)~\cite{hanley1982roc}, which measures discriminative ability across all thresholds, and the Area Under the Precision-Recall Curve (AUPRC), particularly informative under class imbalance~\cite{saito2015precision,mcdermott2024closer}, as anomalous scenes represent a minority of operational traffic.\\

\noindent \textbf{Slice level}: Each 2D axial map is scored independently, and discriminative ability is reported as AUROC, supplemented by two additional operational metrics. The False Positive Rate at 95\% True Positive Rate (FPR@95\% TPR) indicates the false-alarm rate when 95\% of genuine threats are detected. A low FPR@95\% is critical for operational acceptance, as it reflects the proportion of normal cargo subjected to unnecessary secondary inspections. Recall at maximum F1 (Recall@F1) measures the detection rate at the threshold that maximizes the F1-score.

\begin{table*}[htpb!]
\centering
\caption{Slice-level performance metrics across all experimental setups.}
\label{tab:slice_metrics}
\begin{tabular}{c c ccc ccc}
\toprule
\multirow{2}{*}{\textbf{Model}} & \multirow{2}{*}{\textbf{Scoring}} &
\multicolumn{3}{c}{\textbf{IBC1 test}} &
\multicolumn{3}{c}{\textbf{IBC2 test}} \\
\cmidrule(lr){3-5} \cmidrule(lr){6-8}
& & \textbf{AUROC~$\uparrow$} & \textbf{FPR@95\%~$\downarrow$} & \textbf{Recall@F1~$\uparrow$} &
\textbf{AUROC~$\uparrow$} & \textbf{FPR@95\%~$\downarrow$} & \textbf{Recall@F1~$\uparrow$} \\
\midrule
\multirow{3}{*}{IBC1-only}
& MAE     & 0.609 & 0.736 & 0.819 & 0.193 & 0.988 & 1.000 \\
& HI-32   & 0.821 & 0.231 & 0.941 & 0.802 & 0.467 & 0.368 \\
& HI-16   & 0.872 & 0.214 & 0.911 & 0.850 & 0.381 & 0.654 \\
\midrule
\multirow{3}{*}{IBC2-only}
& MAE     & 0.218 & 0.915 & 0.989 & 0.095 & 0.998 & 1.000 \\
& HI-32   & 0.819 & 0.232 & 0.923 & 0.814 & 0.437 & 0.744 \\
& HI-16   & 0.867 & 0.203 & 0.930 & 0.869 & 0.311 & 0.647 \\
\midrule
\multirow{3}{*}{Cold-start}
& MAE     & 0.564 & 0.762 & 0.779 & 0.167 & 0.991 & 1.000 \\
& HI-32   & 0.828 & 0.218 & 0.959 & 0.807 & 0.444 & 0.716 \\
& HI-16   & 0.883 & 0.203 & 0.908 & 0.862 & 0.339 & 0.635 \\
\bottomrule
\end{tabular}

\vspace{0.3cm}
\footnotesize
\textbf{Note:} FPR@95\% = False Positive Rate at 95\% True Positive Rate.
Recall@F1 = Recall at the threshold that maximizes the F1-score.
\end{table*}

\section{Results and Discussion}
\label{sec:results}

This section reports quantitative results at the scene and slice levels, followed by qualitative analyses on synthetic and real scans.

\subsection{Results: Simulation Datasets}

\noindent\textbf{Scene-Level Results}: 
Table~\ref{tab:scene_metrics} reports scene-level AUROC and AUPRC for all model-scoring combinations. In the in-domain setting, pixel-level scoring is highly effective on the synthetic single-IBC configuration: IBC1-only and Cold-start both attain AUROC above $0.98$ with MAE on IBC1 test ($0.983$ and $0.988$, respectively), confirming that the neural network has learned the normality baseline of the cargo configuration. This advantage, however, does not survive the change of cargo configuration. On IBC2, MAE collapses for every model pretrained on IBC1 data, with AUROC dropping to 0.379 (IBC1-only) and 0.429 (Cold-start), that is, below the random-guessing baseline. This is a direct manifestation of domain shift: the global noise floor of IBC2 differs structurally from that of IBC1, and MAE, which averages over all pixels, cannot compensate for it. The HI scores, by contrast, degrade far less across cargo configurations. HI-32 attains at least 0.743 AUROC on IBC2 for all three models, with AUPRC above 0.91 in every case, indicating that cell-based distributional scoring is largely insensitive to the absolute noise floor and therefore resistant to domain shift. The price of this invariance is a moderate loss of in-domain sensitivity on IBC1 (0.897--0.926 AUROC for HI, against 0.983--0.988 for MAE).\\

\noindent The IBC2-only model shows a distinct pattern. Trained exclusively on 35 synthetic IBC2 scenes, it achieves only 0.593 AUROC on IBC2 with MAE and 0.438 on IBC1, reflecting the limitations of a small training set: with roughly a quarter of the scenes of IBC1-only, the reconstruction of the benign background is less faithful (30.14~dB against 31.71~dB validation PSNR, Table~\ref{tab:training}), which increases the residual values uniformly and dominates any pixel-level aggregate. Under spatially aware scoring, this uniform degradation does not affect the score: HI-16 on IBC1 reaches 0.926 AUROC (0.859 AUPRC), on par with IBC1-only (0.922 AUROC, 0.852 AUPRC), a difference smaller than what 20 test scenes can resolve. This indicates that HI does not measure how faithfully the network reconstructs, but whether the reconstruction failure is spatially structured. The Cold-start strategy combines both advantages: initialized from the IBC1-only checkpoint and fine-tuned on the IBC2 training set, it recovers the reconstruction quality of the pretrained model (31.50~dB) while adapting to the dual-container cargo configuration, attaining the best overall balance: 0.988 AUROC on IBC1 with MAE, matching the best HI-32 result on IBC2 (0.750 AUROC, 0.919 AUPRC). We also note that the two HI variants are not uniformly ordered: the finer grid (HI-16) is preferable on IBC1, whereas HI-32 is more reliable on the dual-IBC cargo configuration, where the wider cells average over the structural heterogeneity introduced by the gap between the two containers.\\

\noindent\textbf{Slice-Level Results}: Table~\ref{tab:slice_metrics} complements the scene-level analysis with finer spatial resolution, and exposes the operational limitations of pixel-level scoring in greater detail. MAE exhibits a systematically high Recall@F1 across all settings, indicating that at the threshold maximizing F1, most anomalous slices are flagged. This Recall, however,
comes at a high cost: FPR@95\% exceeds 0.73 for MAE on IBC1 in all models, and approaches 1.00 in every IBC2 evaluation. An FPR@95\% of 0.99 implies that virtually every benign slice would trigger a false alarm at the operating point required to detect 95\% of the threats, which is incompatible with operational conditions in ports. The collapse is most severe across domains: under MAE scoring, the IBC2-only model evaluated on IBC1 reaches only 0.218 AUROC, and the IBC1-only model evaluated on IBC2 only 0.193, both far below chance. The domain shift uniformly raises the global reconstruction noise, and MAE, lacking spatial discriminability, cannot separate this elevated baseline
from the localized weapon signature. HI overcomes this limitation consistently: HI-16 attains an AUROC of at least 0.850 across all six model–configuration combinations, with FPR@95\% below 0.40 in every case. In the in-domain IBC1 setting, Cold-start with HI-16 reaches 0.883 AUROC at an FPR@95\% of 0.203. The IBC2 setting remains more demanding, with HI-16 reaching 0.869 AUROC (IBC2-only) and 0.862 (Cold-start) at FPR@95\% of 0.311 and 0.339, respectively, consistent with the smaller training pool and the greater structural complexity of the dual-IBC cargo configuration.

\begin{figure*}[htbp!]
    \centering
    \begin{subfigure}[t]{0.8\textwidth}
        \centering
        \includegraphics[width=\linewidth]{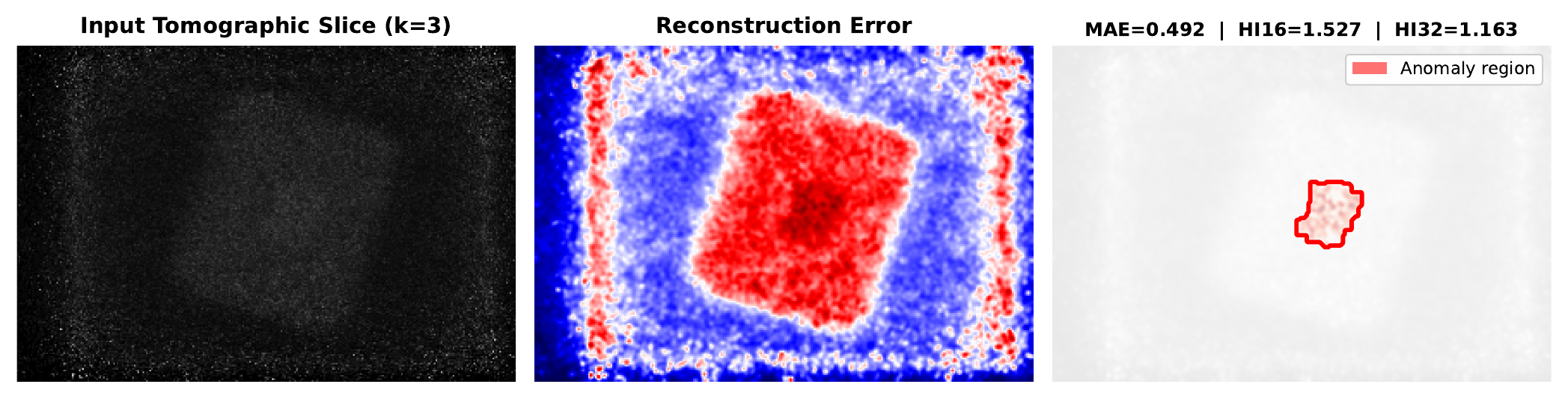}
        \caption{Scene 9 IBC1}
        \label{fig:scene9}
    \end{subfigure}
    \hfill
    \begin{subfigure}[t]{0.8\textwidth}
        \centering
        \includegraphics[width=\linewidth]{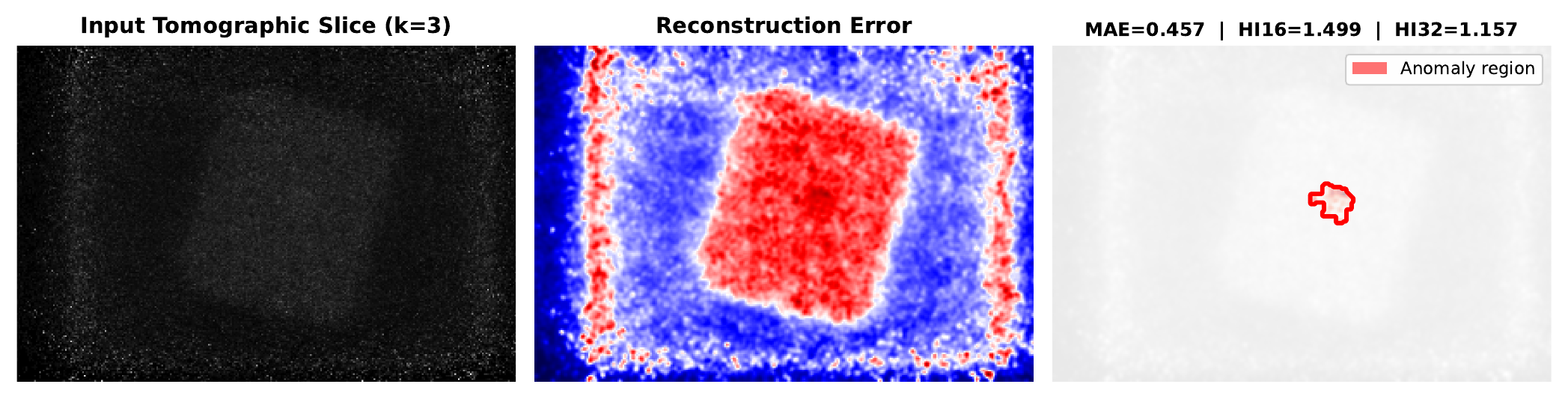}
        \caption{Scene 12 IBC1}
        \label{fig:scene12}
    \end{subfigure}
    \hfill
    \begin{subfigure}[t]{0.8\textwidth}
        \centering
        \includegraphics[width=\linewidth]{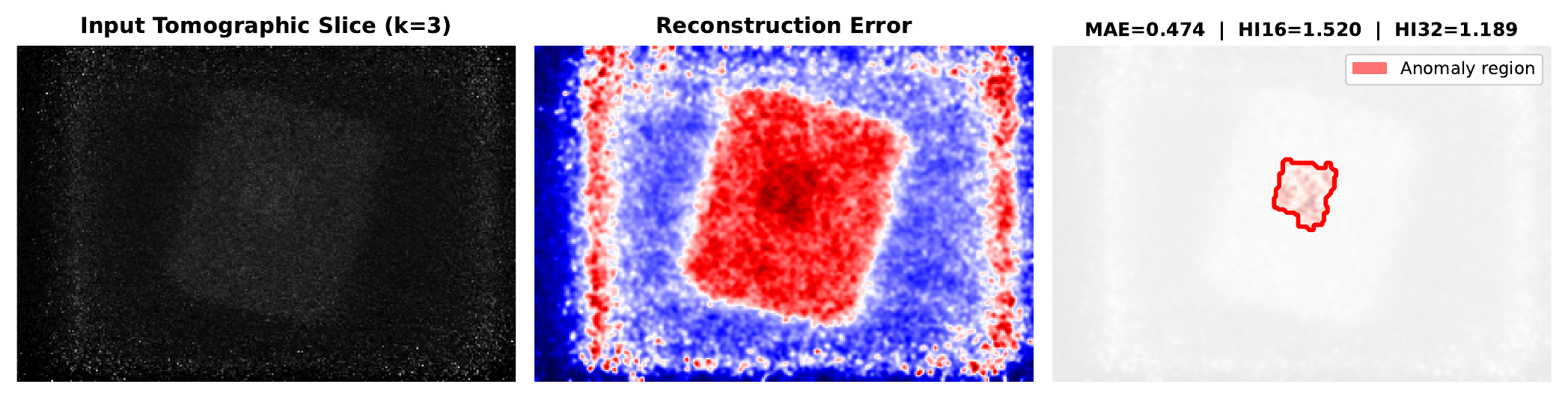}
        \caption{Scene 10 IBC1}
        \label{fig:scene10}
    \end{subfigure}
    \hfill
    \begin{subfigure}[t]{0.8\textwidth}
        \centering
        \includegraphics[width=\linewidth]{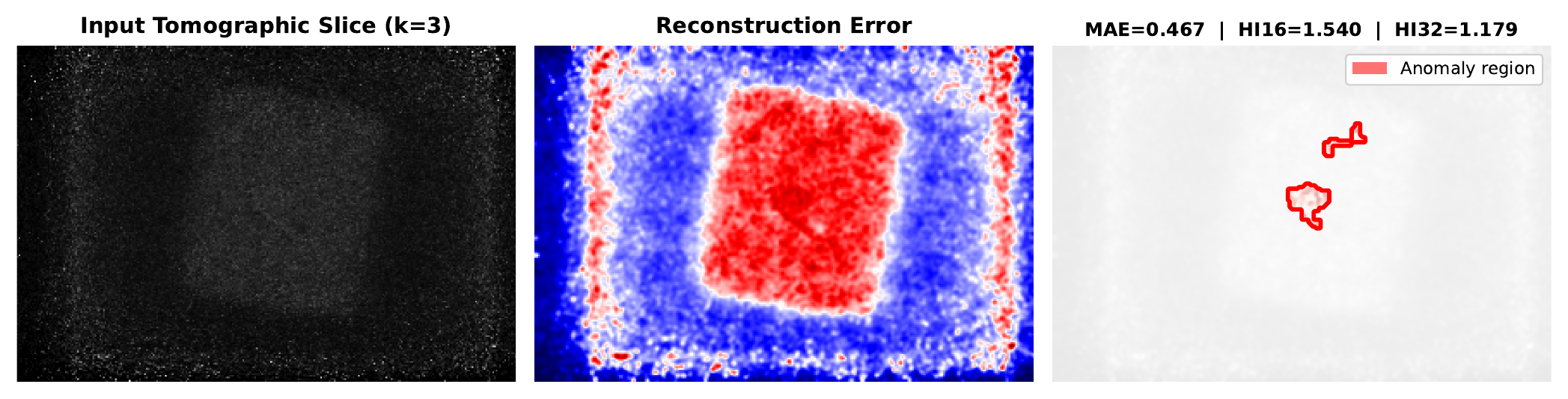}
        \caption{Scene 15 IBC1}
        \label{fig:scene15}
    \end{subfigure}

    \caption{Anomaly detection results for the Cold-start model on the IBC1 synthetic test data. Each panel shows the input 2D scattering map (left), reconstruction (center), and the detected anomaly region along with the scoring metrics (right). Red lines highlight the detection with the anomalous area location.}
    \label{fig:qualitativeIBC1}
\end{figure*}

\begin{figure*}[htbp!]
    \centering
    \begin{subfigure}[t]{0.8\textwidth}
        \centering
        \includegraphics[width=\linewidth]{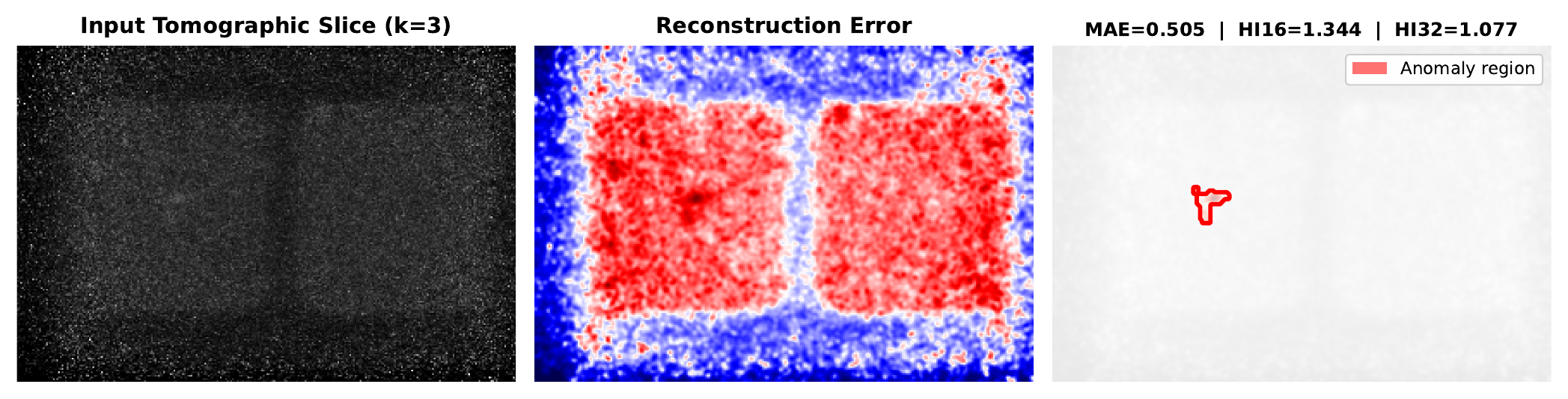}
        \caption{Scene 10 IBC2}
        \label{fig:scene10_IBC2}
    \end{subfigure}
    \hfill
    \begin{subfigure}[t]{0.8\textwidth}
        \centering
        \includegraphics[width=\linewidth]{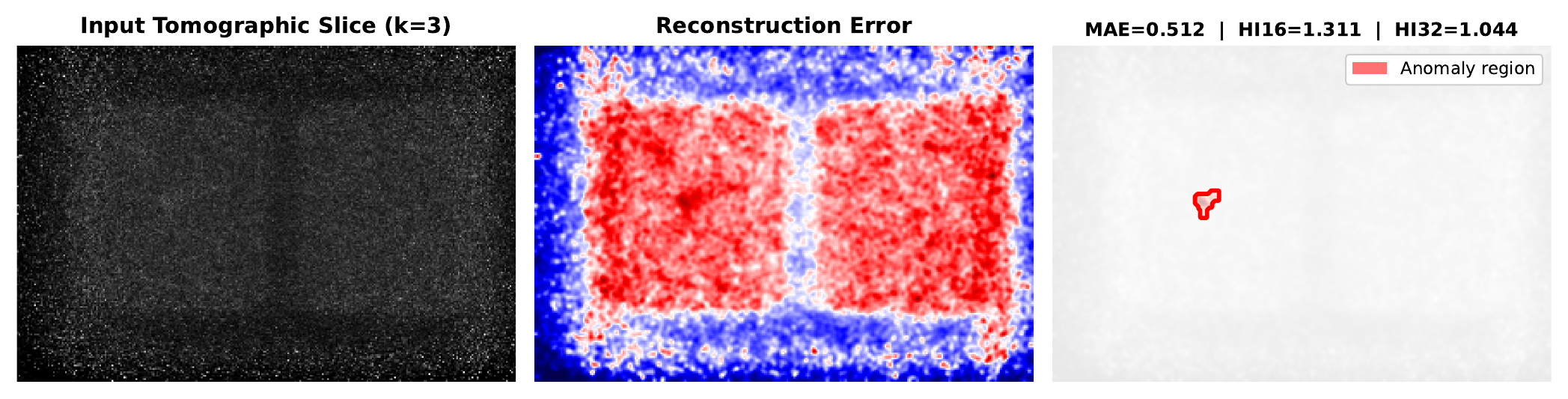}
        \caption{Scene 8 IBC2}
        \label{fig:scene8_IBC2}
    \end{subfigure}
    \hfill
    \begin{subfigure}[t]{0.8\textwidth}
        \centering
        \includegraphics[width=\linewidth]{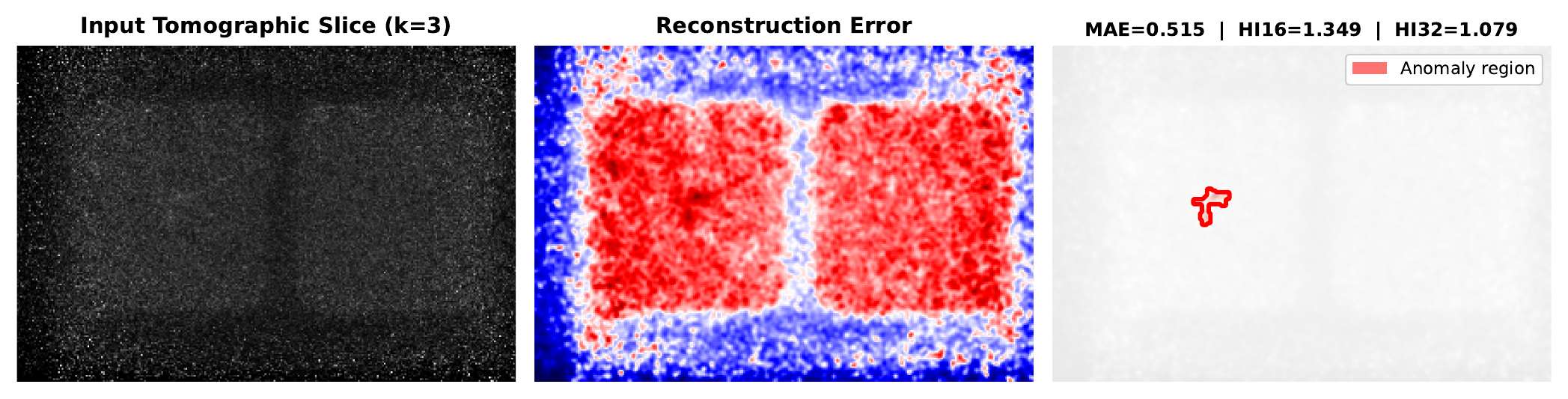}
        \caption{Scene 11 IBC2}
        \label{fig:scene10b_IBC2}
    \end{subfigure}
    \hfill
    \begin{subfigure}[t]{0.8\textwidth}
        \centering
        \includegraphics[width=\linewidth]{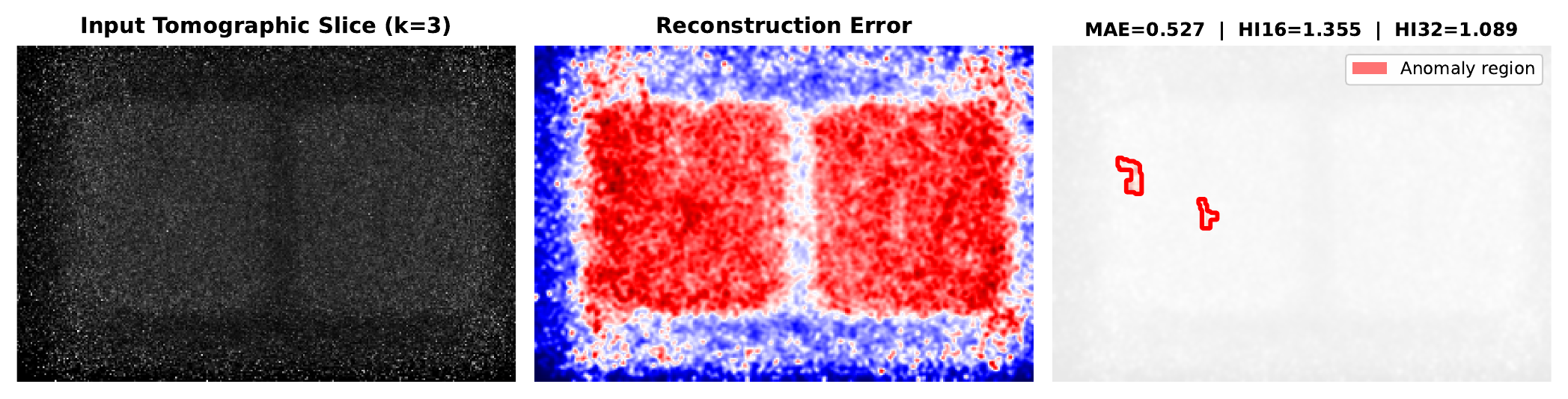}
        \caption{Scene 1 IBC2}
        \label{fig:scene15_IBC2}
    \end{subfigure}

    \caption{Anomaly detection results for the Cold-start model on the IBC2 synthetic test data. Each panel shows the input 2D scattering map (left), reconstruction (center), and the detected anomaly region along with the scoring metrics (right). Red lines highlight the detection with the anomalous area location.}
    \label{fig:qualitativeIBC2}
\end{figure*}

\subsection{Visualization: Simulation Datasets}

Figure~\ref{fig:qualitativeIBC1} presents four representative anomalous slices from the IBC1 synthetic test set. In each panel, the input 2D scattering map (left) displays the characteristic noise floor of a one-hour cosmic-ray acquisition, rendering the cargo structure barely distinguishable from the stochastic background. The reconstruction (center) accurately reproduces the expected benign background but fails to render the anomalous scattering signature, which consequently emerges as a spatially coherent residual in the anomaly map (right). The Cold-start model consistently localizes the weapon region across all four scenes, despite significant variability in depth, lateral position, and liquid composition. The HI-16 and HI-32 scores remain elevated across all scenes, while MAE remains uniformly high due to the spatially uniform noise floor. 

Figure~\ref{fig:qualitativeIBC2} extends the analysis to the dual-IBC configuration. The increased structural complexity, with two separate cargo volumes and a broader reconstruction footprint, introduces a more heterogeneous background, further degrading pixel-level scoring. The HI-based approach nevertheless localizes the weapon within the correct IBC, showing the spatial discriminability of the scoring function across cargo configurations.

In both cargo configurations, the framework produces spurious contours when stochastic noise and reconstruction artifacts are particularly pronounced, as illustrated in Scene 15 (IBC1) and Scene 1 (IBC2) of Figures~\ref{fig:qualitativeIBC1} and~\ref{fig:qualitativeIBC2}. Dead-pixel artifacts and high-magnitude noise spikes produce local error concentrations that the anomaly detection algorithm encloses alongside the true object. This behavior is
a direct consequence of the low muon statistics of one-hour scans and reconstruction artifacts, which leave a residual that never vanishes: the model is trained to flag any deviation from the learned normality, and is not given the means to attribute a deviation to the cargo rather than to the sensor or to the reconstruction. These spurious contours are the main source of false alarms at the slice level and set a practical limit on the precision attainable in the anomaly detection task. The noise floor is intrinsic to the low-statistics regime and cannot be removed: HI only mitigates its effect, by requiring a distributional shift across several cells rather than isolated pixel spikes.

\subsection{Transfer to Real Data}

\begin{figure*}[htbp!]
    \centering
    \begin{subfigure}[t]{0.48\textwidth}
        \centering
        \includegraphics[width=\linewidth]{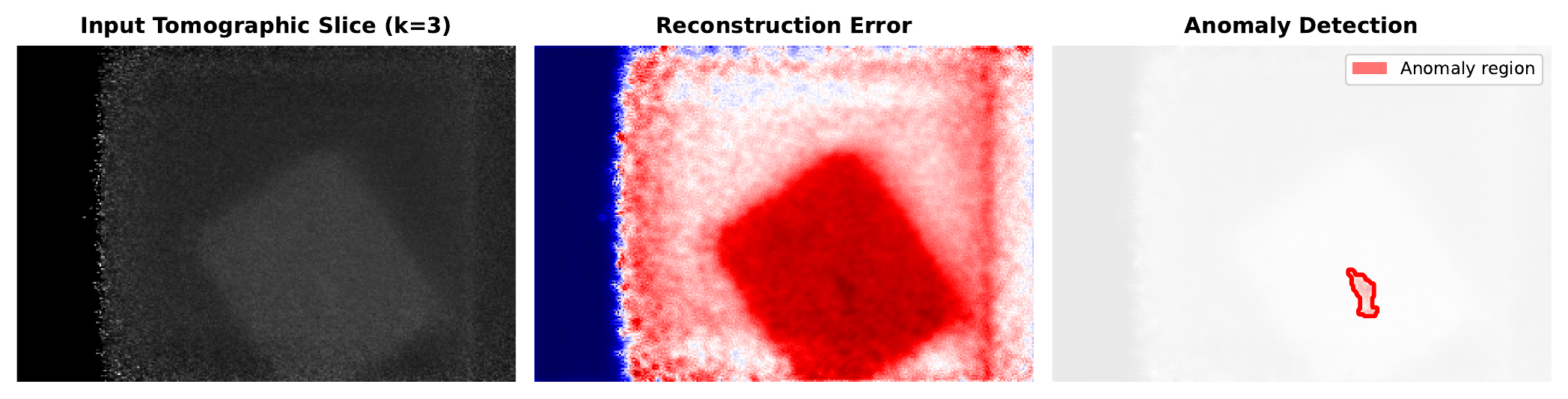}
        \caption{Scene 805 (IBC1)}
        \label{fig:real_1ibc_805}
    \end{subfigure}
    \hfill
    \begin{subfigure}[t]{0.48\textwidth}
        \centering
        \includegraphics[width=\linewidth]{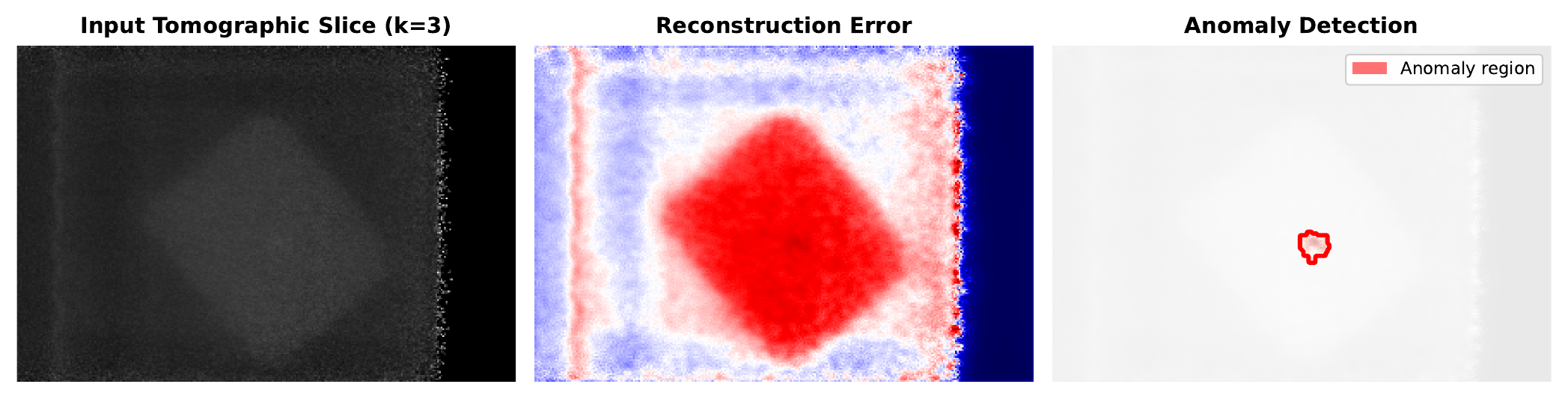}
        \caption{Scene 809 (IBC1)}
        \label{fig:real_1ibc_809}
    \end{subfigure}

    \vspace{0.3cm}

    \begin{subfigure}[t]{0.48\textwidth}
        \centering
        \includegraphics[width=\linewidth]{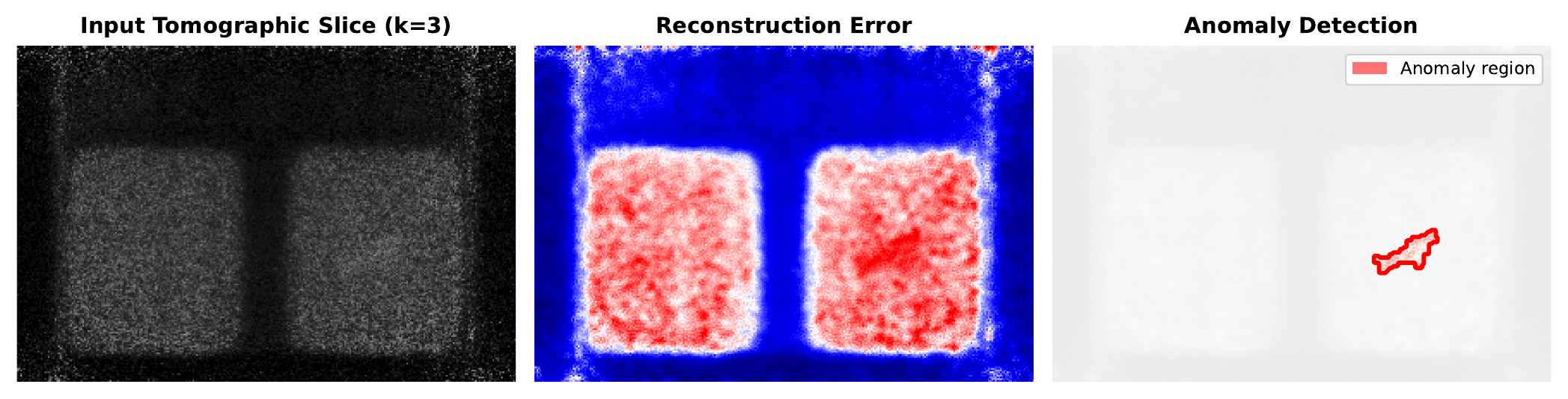}
        \caption{Scene 829 (IBC2)}
        \label{fig:real_2ibc_829}
    \end{subfigure}
    \hfill
    \begin{subfigure}[t]{0.48\textwidth}
        \centering
        \includegraphics[width=\linewidth]{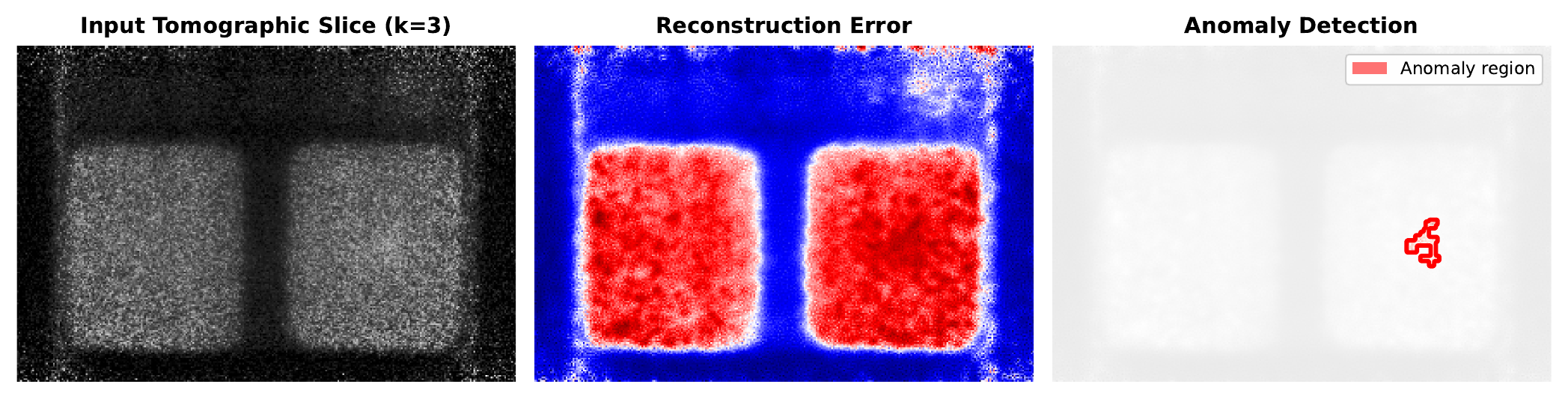}
        \caption{Scene 831 (IBC2)}
        \label{fig:real_2ibc_831}
    \end{subfigure}

    \vspace{0.3cm}

    \begin{subfigure}[t]{0.48\textwidth}
        \centering
        \includegraphics[width=\linewidth]{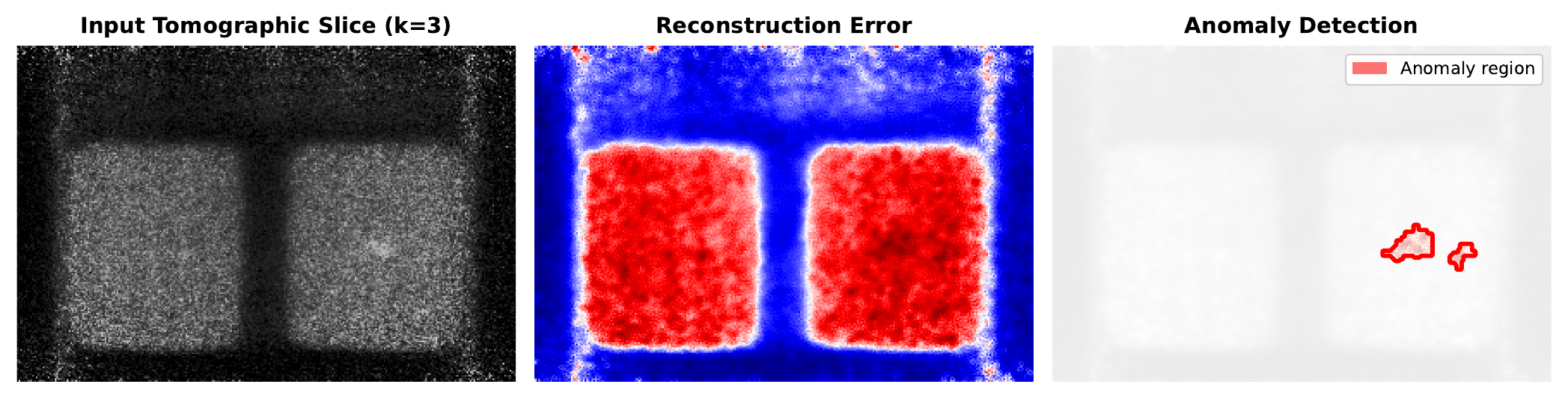}
        \caption{Scene 832 (IBC2)}
        \label{fig:real_2ibc_832}
    \end{subfigure}
    \hfill
    \begin{subfigure}[t]{0.48\textwidth}
        \centering
        \includegraphics[width=\linewidth]{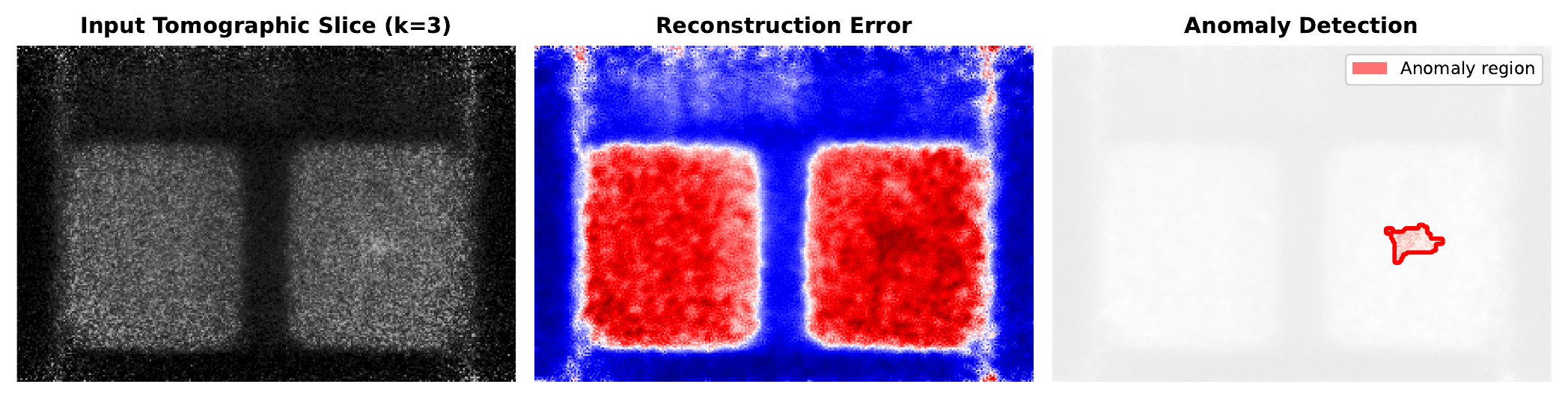}
        \caption{Scene 834 (IBC2)}
        \label{fig:real_2ibc_834}
    \end{subfigure}

    \caption{Anomaly detection results for the Cold-start model on real scan data.
    Panels (a)--(b) correspond to the single-container cargo configuration (IBC1); panels 
    (c)--(f) correspond to the dual-container cargo configuration (IBC2). Each panel shows 
    the input 2D scattering map (left), reconstruction (center), and the detected 
    anomaly region with scoring metrics (right). Red lines highlight the anomalous 
    area location.}
    \label{fig:qualitativeReal}
\end{figure*}

Figure~\ref{fig:qualitativeReal} presents qualitative results on real container scans for both cargo configurations. The IBC1 scans are scenes 805 and 809, both acquired under conditions that introduced geometric misalignments; the IBC2 scans are scenes 829, 831, 832, and 834. Rather than discarding the misaligned scenes, we retain them as a stress test: if the Cold-start model still detects the contraband under suboptimal alignment, this indicates robustness to real-world variability.

In the IBC1 configuration (Scenes 805 and 809), the model reconstructs the benign liquid background faithfully while leaving spatially coherent residuals at the weapon location, which are correctly captured by the HI-based anomaly contour. The IBC1 configuration benefits from a more homogeneous background noise floor, and the anomaly signatures are compact and well localized, even under the geometric misalignment noted above and consistent with the synthetic results of Figure~\ref{fig:qualitativeIBC1}. In the IBC2 configuration (Scenes 829, 831, 832, and 834), the dual-container structure introduces a more complex reconstruction background, with a prominent gap between the two IBCs that generates elevated reconstruction error along the central boundary. Despite this, the HI scoring function discriminates the weapon-induced distributional shift from the structural heterogeneity of the background, localizing the anomaly within the correct IBC in all four scenes. No real measurement entered the training at any point, yet the framework transfers from physically consistent synthetic data to real container scans across two distinct cargo configurations.

\section{Conclusions}
\label{sec:conclusion}

An end-to-end unsupervised anomaly detection framework for maritime MST is presented and validated on real measurements from the SilentBorder demonstration campaign. By framing the problem as an OOD detection task, the framework identifies concealed threats without labeled contraband examples or prior knowledge of threat morphology. The proposed Homogeneity Index scoring function is the component that makes this transferable: while pixel-level metrics remain competitive in-domain on IBC1, they collapse as soon as the cargo configuration changes, whereas HI, by measuring the spatial structure of the reconstruction error rather than its magnitude, is largely insensitive to the noise floor of each acquisition and retains its discriminative power across different cargo configurations. Of the three training strategies evaluated, Cold-start offers the best balance, preserving in-domain sensitivity on the pretraining cargo configuration while matching the best cross-domain performance, which indicates that the sim-to-real gap in MST can be bridged. More broadly, the IBC1-to-IBC2 transition mimics, in a controlled way, the variability encountered in operation, where the cargo configuration changes from one inspection to the next: the ability to adapt from one cargo configuration to another with a small amount of
additional synthetic data, generated once the configuration to be inspected is known, rather than retraining from scratch, makes the approach viable under operational constraints.\\

The results presented here are a first step toward operational anomaly detection in maritime MST with real scan data, and their scope is bounded to the SilentBorder demonstration. Scaling the framework beyond the IBC scenario is the next step: the synthetic dataset, while physically consistent, is tailored to this demonstration setting, and broadening it to a wider variety of concealed objects, liquid compositions, noise regimes, and packaging arrangements is expected to improve both detection and generalization performance across cargo configurations. A further direction is to exploit the geometry of the scene itself: the spatial arrangement of the containers and the cargo are known a priori, and encoding such arrangement, either in the neural network model or in the scoring stage, would allow structural background heterogeneity, such as the gap between two IBCs, to be accounted for rather than competing with the anomaly signal. The modular design of the framework, from the synthetic data pipeline to the HI scoring function, makes these extensions straightforward to incorporate.

\par\vspace{6pt}
\noindent{\bfseries\large Data Statement\par}
\vspace{4pt}
\noindent
Access to representative synthetic data samples may be granted by the corresponding author upon reasonable request. The measured data, collected
before and after the SilentBorder demonstration, cannot be made available.

\par\vspace{6pt}
\noindent{\bfseries\large Conflict of Interest\par}
\vspace{4pt}
\noindent
The authors declare that there is no conflict of interest regarding the publication of this paper.

\par\vspace{6pt}
\noindent{\bfseries\large Use of AI Tools\par}
\vspace{4pt}
\noindent
AI-based tools were used for proofreading and grammar checking.

\par\vspace{6pt}
\noindent{\bfseries\large Acknowledgment\par}
\vspace{4pt}
\noindent

This research was funded by the SilentBorder project under European Union Horizon 2020 grant agreement No 101021812. The authors thank the GScan operational team for carrying out the measurements, and Tarvo Metsalu for arranging the campaigns.


\begin{thebibliography}{99}

\bibitem{TAGAWA2025104376}
H. Tagawa and Q. Meng,
Network evolution of major shipping routes: perspectives from dominant ports and shipping lines,
J. Transp. Geogr. \textbf{128} (2025) 104376

\bibitem{sarah2023}
S. Barnes et al.,
Cosmic-ray tomography for border security,
Instruments \textbf{7} (2023) 13

\bibitem{bonechiyandrea}
L. Bonechi et al.,
Atmospheric muons as an imaging tool,
Rev. Phys. \textbf{5} (2020) 100038

\bibitem{borozdin2003radiographic}
K.N. Borozdin et al.,
Radiographic imaging with cosmic-ray muons,
Nature \textbf{422} (2003) 277

\bibitem{borozdin2023methods}
K. Borozdin et al.,
Muon imaging methods and applications,
Optica Imaging Congress, Technical Digest Series (2023) paper HTu5D.2

\bibitem{odonnell2025upsampling}
W. O'Donnell, D. Mahon, G. Yang and S. Gardner,
Muographic image upsampling with machine learning for built infrastructure applications,
Particles \textbf{8} (2025) 33

\bibitem{riggi2013muon}
S. Riggi et al.,
Muon tomography imaging algorithms for nuclear threat detection inside large volume containers with the Muon Portal detector,
Nucl. Instrum. Methods Phys. Res. A \textbf{728} (2013) 59--68

\bibitem{georgadze2025illicit}
A. Georgadze,
Muon imaging for illicit cargo detection: a simulation-based study,
J. Instrum. \textbf{20} (2025) P06053

\bibitem{sattler2025framework}
F.A. Sattler et al.,
A comprehensive framework toward the seamless integration of muon reconstruction algorithms with machine learning,
J. Appl. Phys. \textbf{138} (2025) 144904

\bibitem{bueno2025pillar}
A. Bueno Rodriguez et al.,
Pillar embedding visualization for muon-scattering tomography,
J. Appl. Phys. \textbf{138} (2025) 14

\bibitem{bury2025momentum}
F. Bury and M. Lagrange,
Scattering-based machine learning algorithms for momentum estimation in muon tomography,
Particles \textbf{8} (2025) 43

\bibitem{medicalref}
J. Bao et al.,
BMAD: benchmarks for medical anomaly detection,
arXiv:2306.11876 (2023)

\bibitem{UNet}
O. Ronneberger, P. Fischer and T. Brox,
U-Net: convolutional networks for biomedical image segmentation,
MICCAI (2015) 234--241

\bibitem{SSIM}
Z. Wang et al.,
Image quality assessment: from error visibility to structural similarity,
IEEE Trans. Image Process. \textbf{13} (2004) 600--612

\bibitem{rodriguez2024b2g4}
A. Bueno Rodriguez et al.,
B2G4: a synthetic data pipeline for the integration of Blender models in Geant4 simulation toolkit,
J. Adv. Instrum. Sci. (2024) 476

\bibitem{eutransportIBC2021}
Western Global,
Intermediate bulk containers --- IBC compliance \& regulations,
Support Insights Guide (2021)

\bibitem{pagano2021ecomug}
D. Pagano et al.,
EcoMug: an efficient cosmic muon generator for cosmic-ray muon applications,
Nucl. Instrum. Methods Phys. Res. A \textbf{1014} (2021) 165732

\bibitem{agostinelli2003geant4}
S. Agostinelli et al.,
GEANT4---a simulation toolkit,
Nucl. Instrum. Methods Phys. Res. A \textbf{506} (2003) 250--303

\bibitem{zaher2025optimization}
Z. Zaher et al.,
Optimization of a cosmic muon tomography scanner for cargo border control inspection,
J. Appl. Phys. \textbf{138} (2025) 19

\bibitem{kiisk2026detector}
M. Kiisk et al., U.S. Patent 12,529,808 (2026)

\bibitem{georgadze2023method}
A. Georgadze et al., U.S. Patent 11,774,626 (2023)

\bibitem{donoho1995}
D.L. Donoho,
De-noising by soft-thresholding,
IEEE Trans. Inf. Theory \textbf{41} (1995) 613--627

\bibitem{zong1998}
X. Zong, A.F. Laine and E.A. Geiser,
Speckle reduction and contrast enhancement of echocardiograms via multiscale nonlinear processing,
IEEE Trans. Med. Imaging \textbf{17} (1998) 532--540

\bibitem{hanley1982roc}
J.A. Hanley and B.J. McNeil,
The meaning and use of the area under a receiver operating characteristic (ROC) curve,
Radiology \textbf{143} (1982) 29--36

\bibitem{saito2015precision}
T. Saito and M. Rehmsmeier,
The precision-recall plot is more informative than the ROC plot when evaluating binary classifiers on imbalanced datasets,
PLOS ONE \textbf{10} (2015) e0118432

\bibitem{mcdermott2024closer}
M.B.A. McDermott et al.,
A closer look at AUROC and AUPRC under class imbalance,
Adv. Neural Inf. Process. Syst. \textbf{37} (2024)

\end{thebibliography}
\end{document}